\documentclass[11pt]{article}

\usepackage{amssymb,latexsym, amsmath}
\usepackage{graphicx}

\makeatletter
\@addtoreset{figure}{section}
\def\thefigure{\thesection.\@arabic\c@figure} \def\fps@figure{h, t}
\@addtoreset{table}{bsection}
\def\thetable{\thesection.\@arabic\c@table} \def\fps@table{h, t}
\@addtoreset{equation}{section}

\makeatother

\advance \topmargin by -\headheight
\advance \topmargin by -\headsep
\evensidemargin \oddsidemargin
\newfont{\tenbi}{cmbxti10}

\begin{document}

\title{On Peakon Solutions of the Shallow Water Equation
}

\author {Mark S. Alber
\thanks{Research partially supported by NSF grant DMS 9626672 and NATO
grant CRG 950897.}\\
Department of Mathematics\\
University of Notre Dame\\
Notre Dame, IN 46556, USA \\
{\footnotesize Mark.S.Alber.1@nd.edu}
\and
Charles Miller\\
Department of Mathematics\\
University of Notre Dame\\
Notre Dame, IN 46556, USA \\
{\footnotesize cmiller6@nd.edu}
}

\date{September 2, 1999}

\maketitle
\begin{abstract}
A new parameterization of the Jacobi inversion problem is used along with
the dynamics of the peaks
to describe finite time interaction of peakon weak solutions of the Shallow
Water
equation.
\end{abstract}

\section{Introduction}
Camassa and Holm [1] described classes of $n$-soliton peaked weak
solutions, or ``peakons," for an integrable (SW) equation
\begin{equation}
\label{CH-sw-eqn}
U_t+3UU_x = U_{xxt}+2U_x U_{xx}+U U_{xxx}-2\kappa U_x\,,
\end{equation}
arising  in
the context of shallow water theory. Of particular interest is their
description of peakon dynamics in terms of a system of completely
integrable Hamiltonian
equations for the locations of the ``peaks'' of the solution, the points at
which its
spatial derivative changes sign. (Peakons have discontinuities
in the
$x$-derivative but both one-sided derivatives exist and differ only by a
sign. This makes
peakons different from
cuspons considered earlier in the  literature.) In other words, each peakon
solution can
be associated with a mechanical system of moving particles.
Calogero [2]
and Calogero and Francoise [3] further extended
the class of mechanical systems of this type.

For the KdV equation,
the spectral parameter $\lambda$ appears linearly in the potential of the
corresponding Schr\"{o}dinger equation: $V = u - \lambda$ in the context of
the
inverse scattering transform (IST) method (see Ablowitz and Segur [4]).
In contrast,
the equation (\ref{CH-sw-eqn}), as
well as $N$-component systems in general, were shown to be connected to
the {\it energy
dependent} Schr\"{o}dinger operators with potentials with poles in the
spectral
parameter.

Alber {\it et al.} [5,6] showed that the presence of a pole
in the potential is essential in a special limiting procedure that
allows for the formation of ``billiard solutions". By using
algebraic-geometric methods, one finds that these billiard
solutions are related to finite dimensional integrable dynamical
systems with reflections. This provides a short-cut to the study of
quasi-periodic and solitonic billiard solutions of nonlinear PDE's.
This method can be used for a number of equations including the
shallow water equation (\ref{CH-sw-eqn}), the Dym type equation, as well as
$N$-component systems with poles and the equations in their
hierarchies [7].  More information on algebraic-geometric methods for integrable
systems can be found in [8] and on billiards in [9,10,11].

In this paper we  consider singular limits of quasi-periodic solutions
when the spectral curve becomes singular and its arithmetic genus drops to
zero.
The solutions are then expressed in terms of purely exponential $\tau$-functions
and they describe the finite time interaction of 2
solitary peakons of the shallow water
equation (\ref{CH-sw-eqn}). Namely, we invert the equations obtained
by using a new parameterization.
First a profile of
the 2-peakon solution is described by considering different
parameterizations for the associated Jacobi inversion problem on three
subintervals of the
$X$-axis and by gluing these pieces of the profile together. The
dynamics of such
solutions is then described by combining these profiles with the dynamics
of the peaks
of the solution in the form developed earlier in Alber {\it et al.} [9,10].
This concludes a derivation in the context of the algebraic
geometric approach of the $n$-peakon ansatz  which was used in the initial
papers [1,12]
for obtaining Hamiltonian systems for peaks. More recently $n$-peakon waves
were studied in [13] and [14].

The problem of describing complex traveling wave and quasi-periodic solutions
of the equation (\ref{CH-sw-eqn})  can be reduced to
solving
finite-dimensional Hamiltonian systems on symmetric products of
hyperelliptic curves.
Namely, according to Alber {\it et al} [5,6,7], such solutions can be
represented in the case of two-phase quasi-periodic   solutions
in the following form
\begin{equation}\label{trace}
U(x,t)=\mu_1+\mu_2 -M,
\end{equation}
where
$M$ is a constant and the evolution of the variables $\mu_1$ and $\mu_2$ is
given by the
equations
\begin{equation}
\label{A-J-g}
\sum_{i=1}^2 \frac{\mu_i^{k}\,{\rm d}\mu_i}{\pm\sqrt{R(\mu_i)}}
= \left\{ \begin{array}{ll}
                            {\rm d} t & \mbox{$k=1,$} \\
                            {\rm d} x & \mbox{$k=2.$} \end{array} \right.
\end{equation}
Here $R(\mu)$ is a polynomial of degree 6 of the form
$R(\mu)=\mu\prod_{i=1}^5(\mu-m_i)$.  The constant from (\ref{trace}) takes
the form $M=1/2\sum m_i$.
Notice that (\ref{A-J-g}) describes quasi-periodic motion on tori of genus
2.  In the limit
$m_1\rightarrow 0$, the solution develops peaks.  (For
details see
Alber and Fedorov [7].)

\noindent
\paragraph{Interaction of Two Peakons.}
In the limit when $m_2\rightarrow m_3\rightarrow a_1$ and
$m_4\rightarrow m_5 \rightarrow a_2$,  we have 2 solitary peakons
interacting with each
other. For this 2 peakon case, we derive the general form of a profile
for a fixed $t$ ($t=t_0,dt=0$) and then see how this profile changes with
time knowing
how the peaks evolve. Notice that the limit depends on the choice of the
branches of the square roots present in (\ref{A-J-g}) meaning choosing
a particular sign
$l_j$ in front of each root. The
problem of finding the profile, after applying the above limits to
(\ref{A-J-g}) gives
\begin{eqnarray}\label{2peaks}
l_1\frac{d\mu_1}{\mu_1(\mu_1-a_1)}+l_2\frac{d\mu_2}{\mu_2(\mu_2-a_1)}&=&
a_2\frac{dX}{\mu_1 \mu_2}=a_2dY \,\\
l_1\frac{d\mu_1}{\mu_1(\mu_1-a_2)}+l_2\frac{d\mu_2}{\mu_2(\mu_2-a_2)}&=&
a_1\frac{dX}{\mu_1 \mu_2}=a_1dY
\end{eqnarray}
where $Y$ is a new variable. This is a new {\it parameterization} of the Jacobi
inversion problem (\ref{A-J-g}) which makes the existence of three
different branches of
the solution obvious.
In general, we consider three different cases: $(l_1=1, l_2=1)$, $(l_1=1,
l_2=-1)$ and
$(l_1=-1,l_2=-1)$. In each case we integrate and invert the integrals to
calculate the symmetric polynomial ($\mu_1 + \mu_2$). After substituting these
expressions into
the trace formula (\ref{trace}) for the solution, this results
in three different parts of the profile   defined on different
subintervals on the real line. The union of these subintervals gives the
whole
line. On the last step these three parts are glued together to obtain a
wave profile with two peaks.

The new parameterization
$dX=\mu_1\mu_2dY$ plays an important role in our approach.  In what follows
each $\mu_i(Y)$
will be defined on the whole real $Y$ line.
However, the transformation from $Y$ back to $X$ is not surjective
so that
$\mu_i(X)$ is only defined on a segment of the real axis.  This is why
different branches are needed to construct a solution on the entire real
$X$ line.

In the case ($l_1=l_2=1$), if we
 assume that there is always one $\mu$
variable between $a_1$ and $a_2$ and one between 0 and $a_1$ and that initial
conditions
are chosen so that
$0<\mu_1^0<a_1<\mu_2^0<a_2$, then we find that:
$
\mu_1+\mu_2=a_1+a_2-(m_1+n_1)a_1a_2 e^{X}.
$
This solution is valid on the domain
$$
X<-\log(a_1n_1+a_2m_1)=X_1^- ,
$$
where $n_1,m_1$ are constants depending on $\mu_1^0,\mu_2^0$.
At the point $X_1^-$,
$$
\mu_1(X_1^-)=0, \hspace{2cm}
\mu_2(X_1^-)=\frac{a_2^2m_1+a_1^2n_1}{a_2m_1+a_1n_1}.
$$
Now we consider
($l_1=-1,l_2=1$).   Here we find the following expression for the symmetric
polynomial
$$
\mu_1+\mu_2=a_1+a_2-\frac{(a_2-a_1)e^{-X}+m_2n_2(a_2-a_1)e^{X}}
{m_2+n_2},
$$
which is only defined on the interval
$$
\log\frac{n_2a_1+m_2a_2}{m_2n_2(a_2-a_1)}>X>
\log\frac{a_2-a_1}{m_2a_1+n_2a_2}=X_1^+.
$$
$m_2,n_2$ are constants which must be chosen so that both $\mu_1$ and $\mu_2$
are continuous at $X_1^-$ and that the ends of the branches match up, that is
so that $X_1^-=X_1^+$.  These conditions are satisfied if
\begin{eqnarray}
m_2&=&\frac{a_2}{a_1}(a_2-a_1)m_1, \,\\
n_2&=&\frac{a_1}{a_2}(a_2-a_1)n_1 . \
\end{eqnarray}
Continuing in this fashion we arrive at the final 3 branched profile
for a fixed $t$,
\begin{eqnarray}\label{sumofmus}
U &=& -(a_1M+a_2N) e^{X} \hspace{5mm}{\rm if } \hspace{2mm} X<-\log(N+M) \,\\
U &=& -\frac{a_1a_2e^{-X}+M\; N\; e^{X}(a_2-a_1)^2}{a_2M+a_1N}\,\\
{\rm if } && \hspace{5mm} -\log(N+M)
<X<\log\frac{a_2^2M+a_1^2N}{(a_2-a_1)^2\;M\;N}\,\\
U &=& -e^{-X}\frac{a_2^3M+a_1^3N}{M\; N(a_2-a_1)^2} \hspace{1cm}
{\rm if } \hspace{2mm} X>\log\frac{a_2^2M+a_1^2N}{(a_2-a_1)^2\; M\; N},  \
\end{eqnarray}
where  we have made the substitution $M=a_2m_1$ and $N=a_1n_1$ and used
the trace formula (\ref{trace}).

{\bf Please place the first figure near here.}

\paragraph{Time evolution.}
So far only a profile has been derived.  Now we will include the time
evolution of the
peaks to find the general solution for the two peakon case.  To
do this we use functions $q_i(t)$ for $i=1,2$ introduced
in Alber {\it et al.} [9]
$$
\mu_i(x=q_i(t),t)=0 ,
$$
for all $t$ and $i=1,2$
which describe the evolution of the peaks. All peaks belong to a zero level set:
$\mu_i=0$. Here the
$\mu$-coordinates, generalized elliptic coordinates,  are used to describe
the positions of the peaks. This yields a connection between $x$ and $t$ along
trajectories of the peaks resulting in a system of equations for the $q_i(t)$.
The solutions of this system are given by
\begin{eqnarray}\label{q}
q_1(t)&=&
q_1^0-a_2t-\log| 1-C_1e^{(a_1-a_2)t} |  +\log(1-C_1) \,\\
q_2(t)&=&
q_2^0-a_2t+\log| 1-C_2e^{(a_2-a_1)t} |  -\log(1-C_2), \label{q2}
\end{eqnarray}
where $C_i=(q_i'(0)-a_1)/(q_i'(0)-a_2)$.

The solution defined in (\ref{sumofmus})  has the peaks given
in terms of the parameters $N$ and $M$.  So to obtain the  solution in
terms of both $x$ and $t$, these parameters must be considered as functions
of time.  The complete solution now has the form
\begin{eqnarray}\label{U}
U &=& -(a_1M(t)+a_2N(t)) e^{X}\, \hspace{5mm}
{\rm if } \hspace{2mm}X<-\log(N(t)+M(t)) \,\\
U &=& \label{U2}
-\frac{a_1a_2e^{-X}+M(t)\; N(t)\; e^{X}(a_2-a_1)^2}{a_2M(t)+a_1N(t)}\,
\nonumber \\
{\rm if } &&
-\log(N(t)+M(t))
<X<\log\frac{a_2^2M(t)+a_1^2N(t)}{(a_2-a_1)^2\;M(t)\;N(t)}\,  \\
U &=& \label{U3}
-e^{-X}\frac{a_2^3M(t)+a_1^3N(t)}{M(t)\; N(t)(a_2-a_1)^2} \, \hspace{5mm}
{\rm if } \hspace{2mm} X >\log\frac{a_2^2M(t)+a_1^2N(t)}{(a_2-a_1)^2\;
M(t)\; N(t)}. \
\end{eqnarray}
where the functions $M(t),N(t)$ are determined by the relations
\begin{eqnarray}
N(t)+M(t)&=&e^{-q_1(t)}=
\frac{ e^{-q_1^0}|e^{a_2t}-C_1e^{a_1t}|}{1-C_1} \,\\
\frac{a_2^2M(t)+a_1^2N(t)}{M(t)\;N(t)}&=&(a_2-a_1)^2e^{q_2(t)}=
 \frac{ (a_2-a_1)^2 e^{q_2^0}|e^{-a_2t}-C_2e^{-a_1t}|}{(1-C_2)} ,
\end{eqnarray}
where $q_1(t), q_2(t)$ are taken from (\ref{q})-(\ref{q2}).  This system
can be solved
to find that
\begin{eqnarray}\label{M}
M(t)&=&\frac{a_1^2 -a_2^2+A(t)B(t)\pm \sqrt{(a_1^2 -a_2^2)^2
-2A(t)B(t)(a_1^2 +a_2^2)+A(t)^2 B(t)^2}}{2B(t)} \,\\
N(t)&=&A(t)-M(t),
\end{eqnarray}
where $A(t)=e^{-q_1(t)}$ and $B(t)=(a_2-a_1)^2e^{q_2(t)}$.
These functions contain 4 parameters, but in fact these can be
reduced to two parameters by using the following relations
\begin{eqnarray}
q_1(0)&=&-\log(M(0)+N(0)) \hspace{2cm}
q_1'(0)=\frac{a_2M(0)+a_1N(0)}{M(0)+N(0)} \,\\
q_2(0)&=&\log\frac{a_2^2M(0)+a_1^2N(0)}
{(a_2-a_1)^2\;M(0)\;N(0)} \hspace{1.1cm}
q_2'(0)=\frac{a_1a_2(a_2M(0)+a_1N(0))}{a_2^2M(0)+a_1^2N(0)} .\
\end{eqnarray}
Some care
must be used in choosing the sign in (\ref{M}).  It is clear that for large
negative $t$, $\mu_1(q_1(t),t)$ refers to the path of one peakon while for
large positive $t$ it refers to the other.  If this were not the case,
simple asymptotic analysis of (\ref{q}) would show that the peakons
change speed which is not the case.  Therefore $q_1(t)$ represents the path
of one of the peakons until some time $t^*$ and the other one after this
time.  The opposite is true for $q_2(t)$.  At the time $t^*$ we say that
a change of identity has taken place.  $t^*$ can be found explicitly by
using the
fact that at this time, the two peaks must have the same height. But the
peaks have the same height exactly when
\begin{equation}\label{t^*}
a_2M(t^*)=a_1N(t^*) .
\end{equation}
Without loss of generality we can rescale time such that $t^*=0$.  In this
case (\ref{t^*}),
due to the original definitions of $m_1,n_1$ given in terms of $\mu_1^0$
$\mu_2^0$,
corresponds to a restriction on the choice of $\mu_1^0$ and $\mu_2^0$, namely
\begin{equation}
-a_2^2\frac{\mu_1^0-a_1}{\mu_1^0-a_2}=a_1^2\frac{\mu_2^0-a_2}{\mu_2^0-a_1} .
\end{equation}
This condition is satisfied for example when
${\displaystyle \mu_1^0=\frac{a_1a_2}{a_1+a_2}}$ and
${\displaystyle \mu_2^0=\frac{a_1+a_2}{2}}$.
Also notice that under this rescaling, the phase shift is simply
$q_1(0)-q_2(0)$.

{\bf Please place the second figure near here}

So we now have a procedure to make  the change of identity occur at
$t=0$, i.e. $\mu_1$ goes from
representing the first peakon to   the second one at $t=0$.  This
change is represented  by the  change in the sign of the
plus/minus in (\ref{M}).  That is, the sign is chosen as positive for
$t<0$ and negative for $t>0$.
However, $M$ remains continuous despite this sign change  since the change of
identity occurs precisely when the term under the square root is zero.
Therefore (\ref{U})-(\ref{U3}) and (\ref{M}) together describe the solution
$U(X,t)$
of the SW
equation as a function of $x$ and $t$
depending on two parameters $M(0)$, $N(0)$.

By using the
approach of this paper  weak billiard solutions can be obtained for
the whole class of $n$-peakon solutions  of $N$-component systems.

\section*{Bibliography.}

\begin{description}

\item [1.] R. Camassa and D. Holm, An integrable
shallow water equation with peaked solitons,
{\it Phys. Rev. Lett.\/}  {\bf 71} 1661-1664 (1993).

\item [2.] F. Calogero, An integrable Hamiltonian system,
{\it Phys. Lett. A.} {\bf 201} 306-310 (1995).

\item [3.] F. Calogero and J. Francoise, Solvable quantum
version of an integrable Hamiltonian system, {\it J. Math. Phys.}
{\bf 37} (6) 2863-2871 (1996).

\item [4.] M. Ablowitz and H. Segur,
Solitons and the Inverse Scattering Transform, SIAM, Philadelphia (1981).

\item [5.] M. Alber, R. Camassa, D. Holm and J. Marsden,
The geometry of peaked solitons and billiard solutions of a class of
integrable PDE's, {\it Lett. Math. Phys.\/} {\bf 32} 137-151 (1994).

\item [6.] M. Alber, R. Camassa, D. Holm, and J. Marsden, On
the link between umbilic geodesics and soliton solutions of nonlinear PDE's,
{\it Proc. Roy. Soc} {\bf 450} 677-692 (1995).

\item [7.] M. Alber and Y. Fedorov, Wave Solutions of Evolution
Equations and Hamiltonian Flows on Nonlinear Subvarieties of Generalized
Jacobians, (subm.) (1999).

\item [8.] E. Belokolos, A. Bobenko, V. Enol'sii, A. Its,  and
V. Matveev, {\it Algebro-Geometric Approach to Nonlinear
Integrable Equations.\/}, Springer-Verlag, Berlin;New York  (1994).

\item [9.] M. Alber, R. Camassa, Y. Fedorov, D. Holm, and J. Marsden,
The geometry of new classes of weak billiard solutions of
nonlinear PDE's. (subm.) (1999).

\item [10.] M. Alber,  R. Camassa, Y. Fedorov, D. Holm and J. Marsden,
On Billiard Solutions of Nonlinear PDE's, {\it Phys. Lett. A} (to appear)
(1999).

\item [11.] Y. Fedorov, Classical integrable systems and billiards
related to generalized Jacobians, {\it Acta Appl. Math.}, {\bf 55} (3)
151--201 (1999).

\item [12.] R. Camassa, D. Holm, and J. Hyman,
A new integrable shallow water equation,
{\it Adv. Appl. Mech.}, {\bf 31} 1--33 (1994).

\item [13.] R. Beals, D. Sattinger, J. Szmigielski, Multipeakons and a
theorem of
Stieltjes, {\it Inverse Problems}, {\bf 15} L1--L4 (1999).

\item [14. ] Y. Li and P. Olver, Convergence of solitary-wave solutions
in a perturbed bi-Hamiltonian dynamical system, {\it Discrete and
continuous dynamical systems}, {\bf 4}, 159--191 (1998).

\end{description}

\newpage

\begin{figure}
\begin{center}
\includegraphics[height=5in]{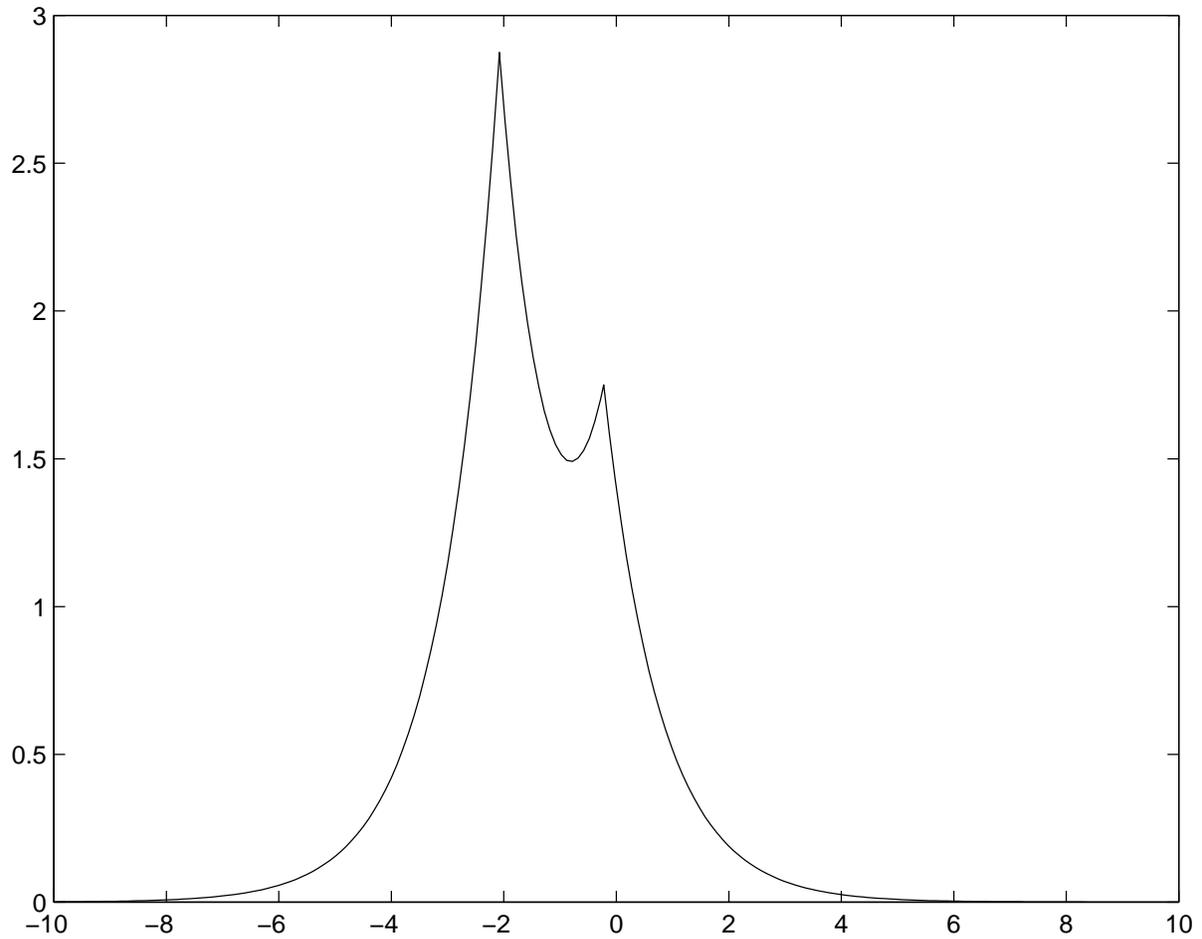}
\end{center}
\caption{\label{fig3}This is a plot $U(x,0)$, a profile of the
 solution to the SW equation
for the two peakon case where $a_1=-1,a_2=-3,\mu_1^0=-.5,\mu_2^0=-1.3.$  Notice
how the solution is defined on three different branches. }
\end{figure}

\newpage

\begin{figure}
\begin{center}
\includegraphics[height=5in]{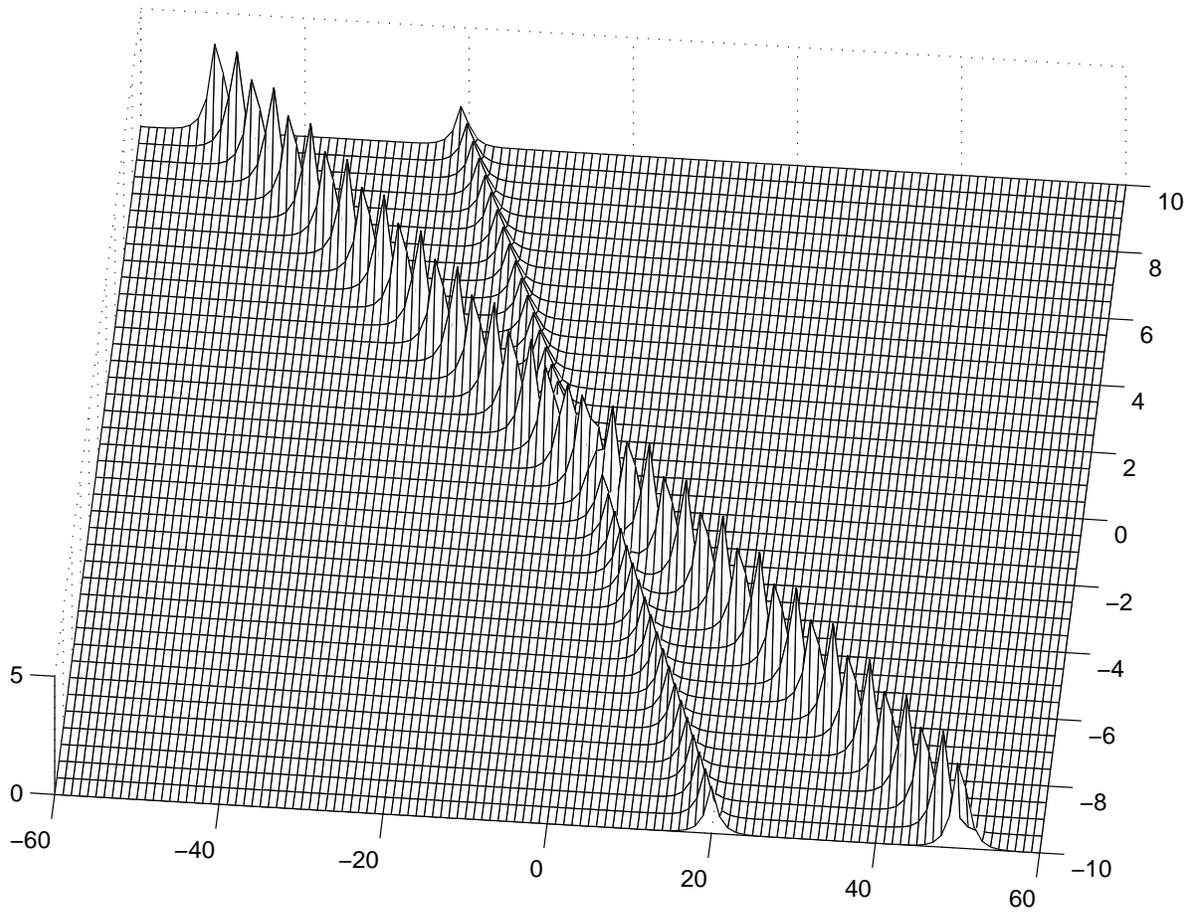}
\end{center}
\caption{\label{fig2}
This is a plot of $(\mu_1+\mu_2)$, which is what we are
seeking in this section.  The parameters used are
$a_1=-1,a_2=-3,\mu_1^0=-.5,\mu_2^0=-1.3$.  Again notice how the solution is
defined
on three different branches. }
\end{figure}

\end{document}